\documentstyle[amssymb,aps,prl,twocolumn,epsfig]{revtex}

\begin{document}

\twocolumn[\hsize\textwidth\columnwidth\hsize\csname
@twocolumnfalse\endcsname

\title{Universal scaling of the Hall resistivity in MgB$_{2}$ superconductors}
\author{W. N. Kang$^{\ast }$, Hyeong-Jin Kim, Eun-Mi Choi,
Hun Jung Kim, Kijoon H. P. Kim, and Sung-Ik Lee}

\address{National Creative Research Initiative Center for Superconductivity
and Department of Physics, Pohang University of Science and Technology,
Pohang 790-784, Republic of Korea }
\draft
\maketitle

\begin{abstract}
The mixed-state Hall resistivity $\rho _{xy}$ and the longitudinal resistivity
$\rho _{xx}$ in superconducting MgB$_{2}$ thin films have been investigated as
a function of the magnetic field over a wide range of current densities from
10$^{2}$ to 10$^{4}$ A/cm$^{2}$. We observe a universal Hall scaling behavior with
a constant exponent $\beta $ of $2.0\pm 0.1$ in
$\rho _{xy}=A\rho _{xx}^{\beta }$, which is independent of the magnetic field,
the temperature, and the current density. This result can be interpreted well within
the context of recent theories.

\end{abstract}

\vskip 1.5pc]

When a type-II superconductor is cooled down from a normal state to a
superconducting state, the Hall effect shows very unusual features, which
have been a long-standing problem and have remained an unresolved issue for
more than three decades. The sign reversal of the Hall effect below T$_{c}$
is one of the most interesting phenomena in the flux dynamics for high-T$_{c}
$ superconductors (HTS) and has attracted both experimental and theoretical
interest. It is now mostly accepted that the mixed-state Hall effect in
type-II superconductors is determined by two contributions, quasiparticle
and hydrodynamic. The sign of the quasiparticle term remains the same as
that in normal state, but the hydrodynamic term can be negative in the mixed
state, so sign reversal can take place \cite
{Dorsey92,Kopnin93,Otterlo95,Ginsberg95}. Furthermore, a scaling behavior
between $\rho _{xy}$ and $\rho _{xx}$ has been found in most HTS \cite
{Samoilov93,Budhani93,Luo92,Kang96,D'Anna00,Kang97,Kang99a,Kang00a}. The
puzzling scaling relation $\rho _{xy}=A\rho _{xx}^{\beta }$ with $\beta \sim
2$ has been observed for Bi$_{2}$Sr$_{2}$CaCu$_{2}$O$_{8}$ crystals \cite
{Samoilov93} and Tl$_{2}$Ba$_{2}$Ca$_{2}$Cu$_{3}$O$_{10}$ films \cite
{Budhani93}. Other similar studies have reported $\beta =1.5\sim 2.0$ for YBa%
$_{2}$Cu$_{3}$O$_{7}$ (YBCO) films \cite{Luo92}, YBCO crystals \cite
{Kang96,D'Anna00}, and HgBa$_{2}$CaCu$_{2}$O$_{6}$ films \cite{Kang97}. Even
$\beta \sim 1$ was reported for heavy-ion-irradiated HgBa$_{2}$CaCu$_{2}$O$%
_{6}$ thin films \cite{Kang99a}.

A number of theories have been proposed to explain the scaling behavior
between $\rho _{xy}$ and $\rho _{xx}$. The first theoretical attempt was
presented by Dorsey and Fisher \cite{Dorsey92b}. They showed that near the
vortex-glass transition, $\rho _{xy}$ and $\rho _{xx}$ could be scaled with
an exponent $\beta =1.7$, and they explained the experimental results of Luo
{\it et al.} for YBCO films \cite{Luo92}. A phenomenological model was put
forward by Vinokur {\it et al.} \cite{Vinokur93}. They claimed that in the
flux-flow region, $\beta $ should be 2 and independent of the pinning
strength. Their result was consistent with the observed exponent in Bi-, Tl-
and Hg-based superconductors only for the Hall data measured in high
magnetic fields\cite{Samoilov93,Budhani93,Kang00a}. Another phenomenological
model was proposed by Wang {\it et al.} \cite{Wang94}, who claimed that $%
\beta $ could change from 2 to 1.5 as the pinning strength increased, which
agreed with the results reported for YBCO crystals \cite{Kang96,D'Anna00}
and Hg-1212 films \cite{Kang97}.

The Hall scaling behavior, therfore, is a complicated phenomena, which seems
strongly depend on the type of HTS. However, from the experimental Hall data
reported in previous papers\cite
{Samoilov93,Budhani93,Luo92,Kang96,D'Anna00,Kang97,Kang99a,Kang00a}, one can
found a general trend. At higher fields, the scaling range is wide and shows
a universal value of $\beta \sim 2$, which is independent of the field and
the pinning strength. At lower fields, the scaling range is relatively
narrow because the contribution from the hydrodynamic term is comparable to
that of quasiparticle term; thus, the value of $\beta $ does not appear to
be constant. Moreover, no Hall scaling has been reported in the field-sweep
data, which provides decisive proof for the temperature dependence of the
Hall scaling behavior.

The MgB$_{2}$ superconductor is a very interesting sample for investigating
the flux dynamics. Different from HTS, MgB$_{2}$ shows no Hall sign anomaly
in\ the mixed state and has a rather simple vortex phase diagram\cite
{Kang01b}. The absence of sign anomaly implies that the hydrodynamic
contribution is very small or negligible. Thus, the MgB$_{2}$ compound is
probably the best candidate for probing whether the Hall scaling is
universal or not because we need only consider the quasiparticle term of the
Hall conductivity, which is consistent with the universal Hall scaling theory
\cite{Vinokur93}.

In this Letter, we report the first demonstration of a universal scaling
behavior of the Hall resistivity in MgB$_{2}$ superconducting thin films,
and the results can be well described using recent theories. We confirmed
that the scaling exponent $\beta \sim 2$ is universal, which is independent
of the temperature, the magnetic field, and the pinning strength. In order
to test the pinning strength dependence of the scaling behavior, we measured
the Hall effect for two orders of magnitude of the current density. Based on
our results, we will show that the universal Hall scaling law is also valid
for HTS in high fields where the hydrodynamic contribution in the Hall
effect is very small compared to the quasiparticle contribution.

The MgB$_{2}$ thin films were grown on Al$_{2}$O$_{3}$ (1 $\bar{1}$ 0 2)
substrates under a high-vacuum condition of $\sim $ 10$^{-7}$ Torr by using
pulsed laser deposition and postannealing techniques. The fabrication
process and the normal-state transport properties of MgB$_{2}$ thin films
are described in detail elsewhere\cite{Kang01b,Kang01c}. The X-ray
diffraction patterns indicated highly c-axis-oriented thin films
perpendicular to the substrate surface and with a sample purity in excess of
99$\%$. The critical current density at 15 K and under a self-field
condition was observed to be on the order of 10$^{7}$ A/cm$^{2}$ \cite
{HJKim01}. Standard photolithographic techniques were used to produce
thin-film Hall bar patterns, which consisted of a rectangular strip (1 mm $%
\times $ 3 mm) of MgB$_{2}$ film with three pairs of sidearms. The narrow
sidearm width of 0.1 mm was patterned so that the sidearms would have an
insignificant effect on the equipotential. Using this 6-probe configuration,
we were able to measure simultaneously the $\rho _{xx}$ and the $\rho _{xy}$
at the same temperature. To achieve good ohmic contacts, we coated Au film
on the contact pads after cleaning the sample surface by using Ar-ion
milling. Fine temperature control was crucial since the Hall signal of MgB$%
_{2}$ is very small. The magnetic field was applied perpendicular to the
sample surface by using a superconducting magnet system, and the applied
current densities were 10$^{2}-$ 10$^{4}$ A/cm$^{2}$. The Hall voltage was
found to be linear in both the current and the magnetic field.

Figure 1(a) shows the temperature dependence of $\rho _{xx}$ for MgB$_{2}$
thin films for various magnetic fields up to 5 T and for applied current
densities of 10$^{3}$ and 10$^{4}$ A/cm$^{2}$. At zero field, the onset
transition temperature ($T_{c}$) was 39.2 K and had a narrow transition
width of $\sim $ 0.15 K, as judged from the 10 to 90$\%$ superconducting
transition. As current density was decreased, the large enhancement of \ T$%
_{c}$ was clearly observed. This is similar pinning effect being enhanced by
introducing columnar defects in HTS\cite{Budhani93,Kang96,Kang99a}. This
result indicates that we can investigate the pinning strength dependence of
the Hall scaling behavior by changing the applied current density.

The corresponding $\rho _{xy}$ is plotted in Fig. 1(b). The same trend as
seen in the $\rho _{xx}-T$ curves was observed with increasing current
density. No sign change was detected for magnetic fields up to 5 T over a
wide range of current densities from 10$^{3}$ to 10$^{4}$ A/cm$^{2}$. This
result implies that in the MgB$_{2}$ superconductor, the hydrodynamic
contribution for the mixed-state Hall conductivity is very small or
negligible compared to the quasiparticle contribution. Thus, this compound
is probably the most suitable sample to prove the Hall scaling behavior. The
overall feature of the temperature dependence is quite different from that
of HTS \cite{Samoilov93,Budhani93,Kang96,Kang97,Kang99a,Kang00a}, in which a
dip structure is observed near T$_{c}$ due to the negative contribution by
the hydrodynamic term. The normal-state $\rho _{xy}$ at 5 T decreases
linearly with increasing temperature from 30 to 40 K, which is consistent
with the temperature dependence of $\rho _{xy}$ above T$_{c}$\cite{Kang01c}.

Figure 2 shows the field dependence of (a) $\rho _{xx}$ and (b) $\rho _{xy}$
for MgB$_{2}$ films at various temperatures from 28 to 40 K and current
densities of 10$^{3}$ and 10$^{4}$ A/cm$^{2}$. A very small and positive
magnetoresistance was observed at 40 K and 5 T. With decreasing temperature
and current density, the superconducting transitions of $\rho _{xx}$ and $%
\rho _{xy}$ became broad, showing that a relatively wide vortex-liquid phase
had been formed. Similar to the $\rho _{xx}-T$ behavior, large increases of
zero-resistance transition were observed in both the $\rho _{xx}$ and the $%
\rho _{xy}$ data when a lower current density was applied, which indicates
that the pinning force can be adjusted systematically by changing the
applied current density. As the magnetic field was increased, $\rho _{xy}$
grew gradually without changing its sign up to an upper critical field
(marked by the arrow), which is different from the previous observation\cite
{Jin01} in polycrystalline MgB$_{2}$. In the normal state above upper
critical fields, we found a quite linear $H$-dependence of $\rho _{xy}$.
This is clearly different from the HTS case and suggests a relatively simple
flux dynamics in MgB$_{2}$.

The scaling behavior of $\rho _{xy}$ between $\rho _{xx}$ for the
temperature-sweep data (Fig. 1) of MgB$_{2}$ films is plotted in Fig. 3 for
various fields from 1 to 5 T and over a wide range of current densities from
10$^{2}$ to 10$^{4}$ A/cm$^{2}$. The universal Hall scaling with an exponent
of 2.0 $\pm $ 0.1 is evident and is independent of the magnetic field and
the current density. More strikingly, this universal scaling generally
occurs in the regimes of the flux flow, such as the free-flux-flow, the
thermally activated flux-flow, and the vortex-glass regions\cite{HJKim01b}.
This result is different from those for HTS where the scaling relation is
valid only in the thermally activated flux-flow and vortex-glass regions\cite
{Samoilov93,Budhani93,Luo92,Kang96,D'Anna00,Kang97,Kang99a,Kang00a}.

Other supporting data for the universal Hall scaling of the field-sweep data
(Fig. 2) can be seen in Fig. 4. We find a universal value of $\beta $ = 2.0 $%
\pm $ 0.1, which does not depend on the temperature or on the current
density at \ (a) 10$^{4}$ and (b) 10$^{3}$ A/cm$^{2}$. This is the first
observation using the field-sweep measurements and provides decisive
evidence for the temperature independence of the Hall scaling. Thus, we have
confirmed that the flux-flow $\rho _{xy}$ is determined by $\rho _{xx}$ and
that the relation is independent of the magnetic field, the temperature, and
the current density. This universal behavior of the Hall scaling is our
principal finding, and this observation will have serious implications for
the physics of mixed-state Hall behavior, as discussed below.

Vinokur {\it et al}.\cite{Vinokur93} proposed a universal Hall scaling
theory based on a force balance equation where the Lorentz force ${\bf f}_{L}
$ acting on a vortex is balanced by the usual frictional force $-\eta ${\bf v%
} and the Hall force $\alpha ${\bf v} $\times $ {\bf n }with {\bf v} being
the average velocity of vortices and {\bf n} being the unit vector along the
vortex lines. The coefficient $\alpha $ is related to the Hall angle by
means of tan$\Theta _{H}=\alpha /\eta $. If a pinning force of $-\gamma $%
{\bf v}, which renormalizes only the frictional term but does not affect the
Hall force term, is included, the equation of vortex motion is given by

\begin{equation}
(\gamma +\eta ){\bf v+}\alpha {\bf v}\times {\bf n}={\bf f}_{L}\text{.}
\end{equation}
Using this force balance equation, we can easily calculate the relation
between $\rho _{xy}$ and $\rho _{xx}$:

\begin{equation}
\rho _{xy}=A\rho _{xx}^{2}\text{,}
\end{equation}
where $A=\alpha /(\Phi _{0}B)$ with $\Phi _{0}$ being the flux quantum.
Equation (2) gives a universal scaling law with $\beta =2$, which is
independent of the magnetic field, the temperature, and the pinning
strength. Furthermore, this relation can applied to the entire flux-flow
region. Our experimental data in Figs. (3) and (4) can be interpreted
completely by this simple theory. This excellent consistency between
theoretical and experimental results is very important in the field of
vortex dynamics since we are able to set up the basic equation of vortex
motion, Eq. (1), which should provide a significant direction for future
investigations on vortex dynamics. If this theory is to be generally
accepted, however, experimental observation of HTS with $\beta $ ranging
from 1 to 2 should be explained.

Now, we discuss the spread value of $\beta $ ($1<\beta <2$) reported for
HTS. An interesting microscopic approach based on the time-dependent
Ginzburg-Landau theory has been proposed in a number of papers \cite
{Dorsey92,Kopnin93,Otterlo95}. According to this model, the mixed-state Hall
conductivity $\sigma _{xy}$ in type II superconductors is determined by the
quasiparticle contribution $\sigma _{xy}^{(q)}$ and the hydrodynamic
contribution $\sigma _{xy}^{(h)}$ of the vortex cores, $\sigma _{xy}=\sigma
_{xy}^{(q)}+\sigma _{xy}^{(h)}$. Since $\sigma _{xy}^{(h)}$ is determined by
the energy derivative of the density of states \cite{Kopnin93,Otterlo95}, if
that term is negative and dominates over $\sigma _{xy}^{(q)}$, a sign
anomaly can appear. This theory is consistent with experimental data for HTS
\cite{Ginsberg95}. This microscopic theory suggests that Hall scaling can be
broken in the case where $\sigma _{xy}^{(h)}$ is comparable to $\sigma
_{xy}^{(q)}$ because those terms have opposite signs. Indeed, $\beta $ was
observed to less than $2$ at low fields whereas a universal value of $\beta
\thicksim 2$ was observed at higher fields where $\sigma _{xy}^{(h)}$ is
very small compared to $\sigma _{xy}^{(q)}$ (for example, $\beta $ $%
\thicksim 2$ was observed in Hg- and Tl-based superconductors for $H=9-18$ T
\cite{Kang00a}). For the MgB$_{2}$ compound, since no sign anomaly was
detected, we can say with confidence that $\sigma _{xy}^{(h)}$ is very small
or negligible; thus, universal Hall scaling holds over a wide range of
fields. One can conclude from these results, that Hall scaling is universal
under conditions where the $\sigma _{xy}^{(q)}$ term dominates the $\sigma
_{xy}^{(h)}$ term; thus we can then explain all the reported data related
Hall scaling issues for HTS \cite
{Samoilov93,Budhani93,Luo92,Kang96,D'Anna00,Kang97,Kang99a,Kang00a}.

In summary, we have found a universal Hall scaling behavior between $\rho
_{xy}$ and $\rho _{xx}$ in MgB$_{2}$ thin films, which is in good agreement
with the high-field Hall data from Bi-, Hg- and Tl-based HTS\cite
{Samoilov93,Kang00a}. Our Hall data can be completely explained within the
context of the universal Hall scaling theory. We also show that $\rho _{xy}$
can scale as $\rho _{xx}^{2}$ in cases where the $\sigma _{xy}^{(q)}$ term
dominates $\sigma _{xy}^{(h)}$ term. With a simple phenomenological theory
\cite{Vinokur93} and a microscopic theory\cite{Dorsey92,Kopnin93,Otterlo95},
we are able to explain the Hall scaling behavior in HTS, which has been
debated for a long time. We believe that these results will provide new
insight into the future theoretical studies on the vortex dynamics of
superconductivity.

This work is supported by the Creative Research Initiatives of the Korean
Ministry of Science and Technology.

\bigskip

\begin{figure}[tbp]
\centering \epsfig{file=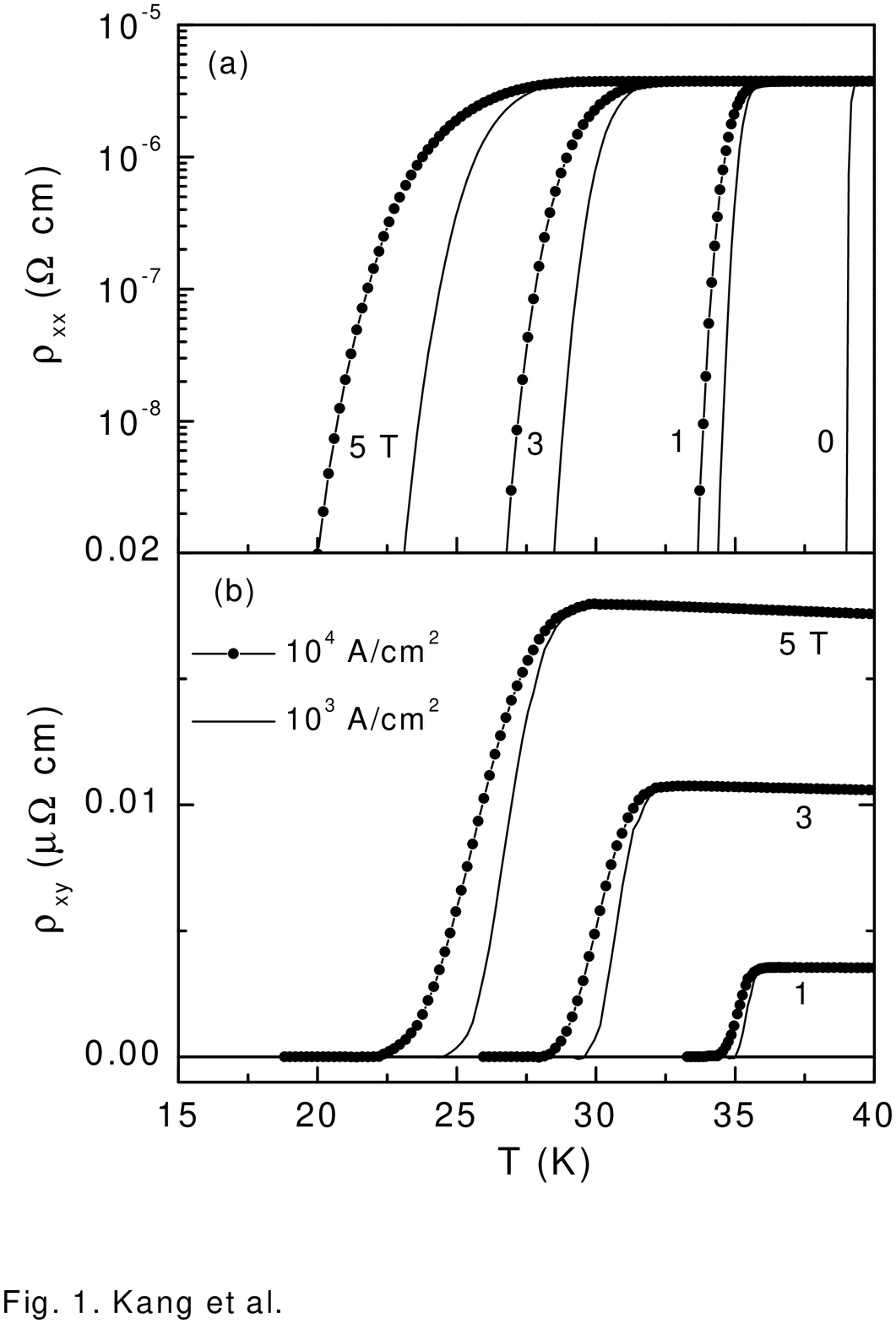, width=7.5cm}
\caption{Temperature dependences of (a) $\protect\rho _{xx}$ and (b) $%
\protect\rho _{xy}$ for MgB$_{2}$ thin films in a magnetic field up to 5 T
and at current densities of 10$^{3}$ and 10$^{4}$ A/cm$^{2}$.}
\end{figure}

\begin{figure}[tbp]
\centering \epsfig{file=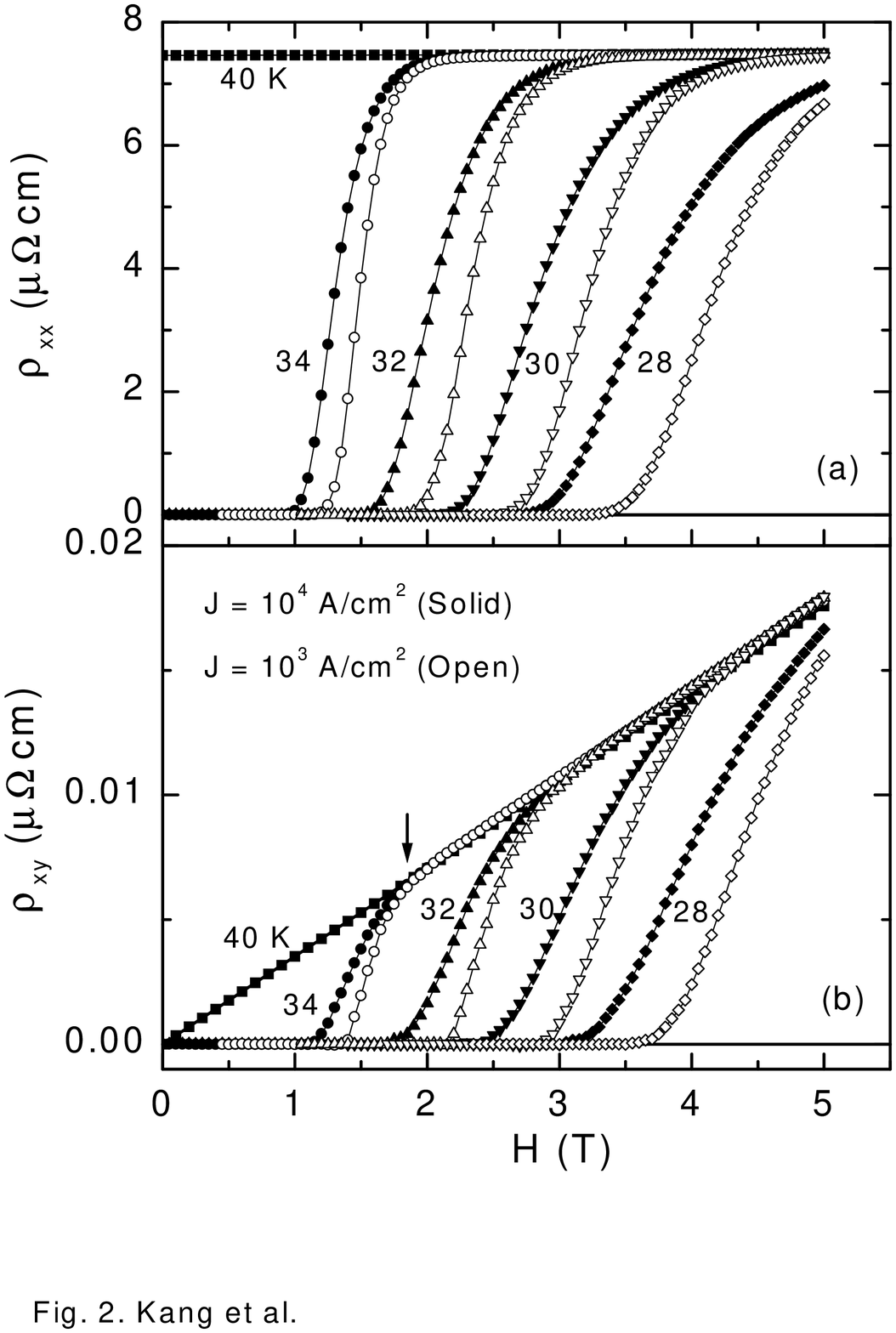, width=7.5cm}
\caption{Magnetic field dependences of (a) $\protect\rho _{xx}$ and (b) $%
\protect\rho _{xy}$ curves for MgB$_{2}$ thin films at temperatures from 28
to 40 K and current densities of 10$^{3}$ and 10$^{4}$ A/cm$^{2}$.}
\end{figure}
\begin{figure}[tbp]
\centering \epsfig{file=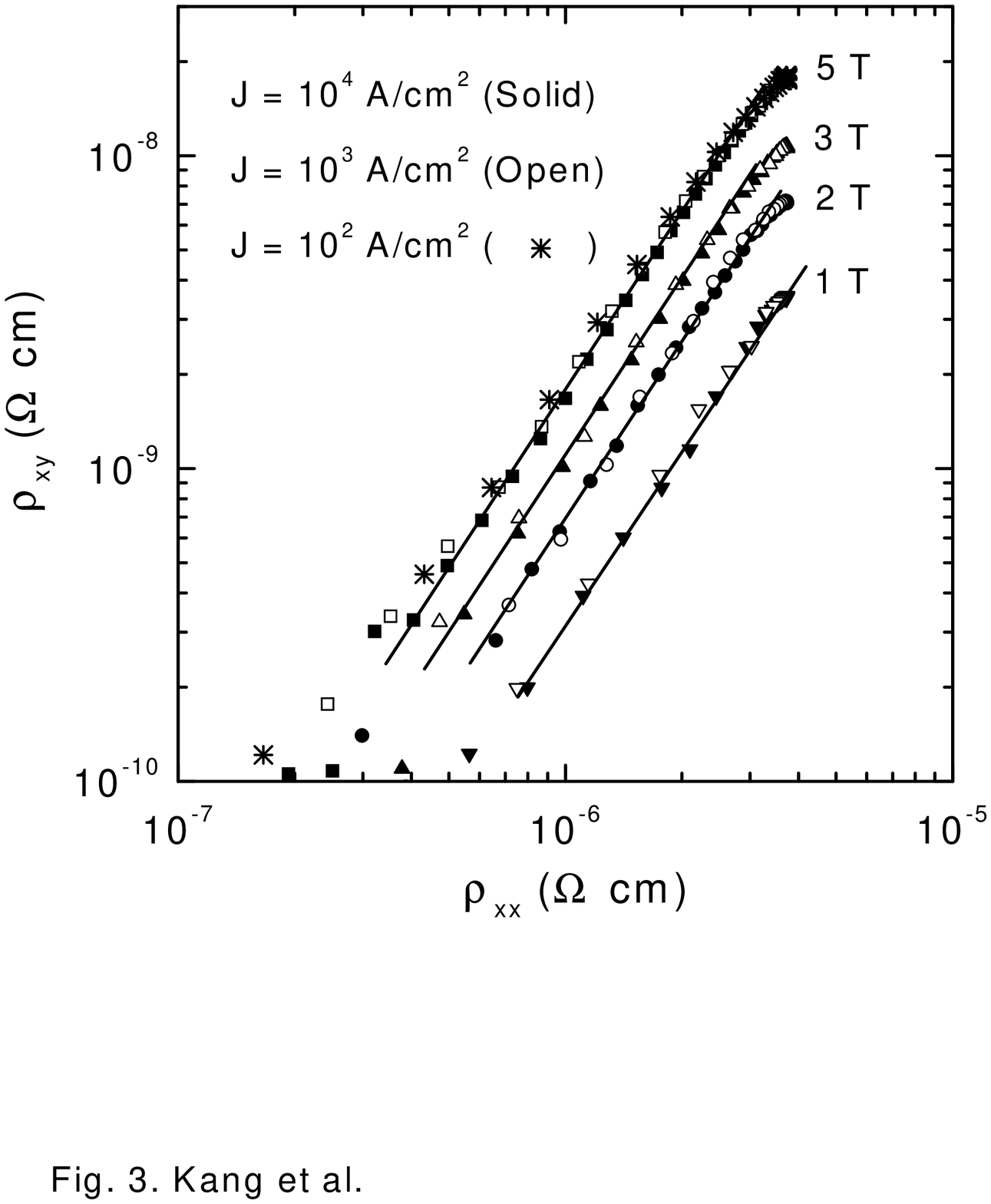, width=7.5cm}
\caption{Scaling behaviors between $\protect\rho _{xy}$ and $\protect\rho %
_{xx}$ for the temperature-sweep data of MgB$_{2}$ thin films in magnetic
fields from 1 to 5 T and at current densities of 10$^{2}$, 10$^{3}$, and 10$%
^{4}$ A/cm$^{2}$. The universal Hall scaling with an exponent of
2.0 $\pm $ 0.1, which is independent of magnetic fields and
current densities, is evident.}
\end{figure}
\begin{figure}[tbp]
\centering \epsfig{file=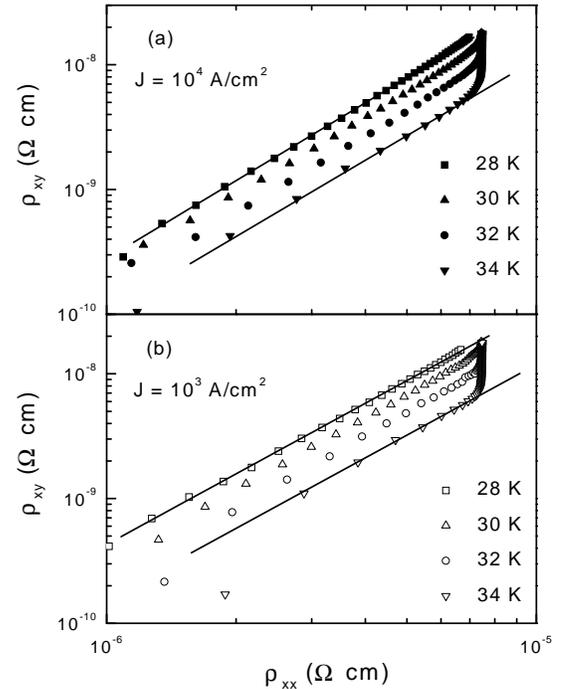, width=7.5cm}
\caption{Scaling behaviors between $\protect\rho _{xy}$ and $\protect\rho %
_{xx}$ for the field-sweep data of MgB$_{2}$ thin films measured at
temperatures from 28 to 34 K and current densities of (a) 10$^{4}$ and (b) 10%
$^{3}$ A/cm$^{2}$. A universal Hall scaling with an exponent of 2.0 $\pm $
0.1, which is independent of the temperature and the current density, is
clearly seen.}
\end{figure}

\end{document}